\documentclass[a4paper,fleqn,usenatbib]{mnras}

\usepackage{newtxtext,newtxmath}

\usepackage[T1]{fontenc}
\usepackage{ae,aecompl}

\usepackage{graphicx}  
\usepackage{amsmath}   
\usepackage{amssymb}   

\newcommand{\ccm}{\,\text{cm}^{-3}}

\newcommand{\Kccm}{\,\text{K cm}^{-3}}

\newcommand{\NH}{\,N$_2\text{H}^+$}

\title[Morphology of prestellar cores in filaments]{Morphology of prestellar cores in pressure confined filaments}

\author[S. Heigl et al.]{
S. Heigl,$^{1,2}$\thanks{E-mail: heigl@usm.lmu.de}
M. Gritschneder,$^{1}$
A. Burkert,$^{1,2}$
\\
$^{1}$Universit\"ats-Sternwarte, Ludwig-Maximilians-Universit\"at M\"unchen, Scheinerstr. 1, 81679 Munich, Germany\\
$^{2}$Max-Planck Institute for Extraterrestrial Physics, Giessenbachstr. 1, 85748 Garching, Germany\\
}

\date{Accepted XXX. Received YYY; in original form ZZZ}

\pubyear{2018}

\begin{document}
\label{firstpage}
\pagerange{\pageref{firstpage}--\pageref{lastpage}}
\maketitle

\begin{abstract}
   Observations of prestellar cores in star-forming filaments show two distinct
   morphologies. While molecular line measurements often show broad
   cores, submillimeter continuum observations predominantly display pinched
   cores compared to the bulk of the filament gas. In order to explain
   how different morphologies arise, we use the gravitational
   instability model where prestellar cores form by growing density
   perturbations. The radial extent at each position is set by the local
   line-mass. We show that the ratio of core radius
   to filament radius is determined by the initial line-mass of the
   filament. Additionally, the core morphology is independent of perturbation
   length scale and inclination, which makes it an ideal diagnostic for
   observations. Filaments with a line-mass of less than half its critical
   value should form broad cores, whereas filaments with more than half its
   critical line-mass value should form pinched cores. For filaments embedded
   in a constant background pressure, the dominant perturbation growth times
   significantly differ for low and high line-mass filaments. Therefore, we
   predict that only one population of cores is present if all filaments within
   a region begin with similar initial perturbations.
\end{abstract}




\section{Introduction}
\label{sec:introduction}

   It has long been proposed that core formation in filaments is tied to some
   kind of fragmentation process \citep{schneider1979, larson1985}.
   This connection has only been reinforced by observations of
   the \textit{Herschel} Space Observatory \citep{andre2010,
   koenyves2010, menshchikov2010, ward-thompson2010,
   arzoumanian2011, arzoumanian2013, kirk2013, andre2014}, which show
   that dense cores are contained in an ubiquitous filamentary structure in
   molecular clouds. As cores are the birth-site of stars \citep{benson1989,
   klessen1998, mckee2007}, it is essential to understand the process of core
   formation in order to develop a coherent model for stellar formation.
   Different models of core formation have been proposed, e.g. by the
   dissipation of turbulence \citep{padoan2001, klessen2005} or by collapse
   of density enhancements due to intersecting filaments, so called "hubs"
   \citep{myers2009}. The complexity of core formation has increased with the
   observations of fibres \citep{hacar2013, tafalla2015}, trans- and subsonic
   velocity coherent substructures in filaments, again opening the possibility
   that cores form by subsonic motions due to gravitational instabilities,
   potentially modified by magnetic fields either hindering core formation
   due to magnetic pressure \citep{nagasawa1987, gehman1996b, fiege2000} or
   facilitating core formation in a magnetically stabilized filament by
   ambipolar diffusion \citep{shu1987, hosseinirad2017}.\\
   A possible indicator to validate this model is the comparison of observed
   cores with the analytical predictions of overdensities forming by
   gravitational instabilities. High dynamic range observations in the
   submillimeter continuum, for instance in the Taurus region, show very thin
   cores compared to the filament radius \citep{marsh2014}. Contrarily,
   molecular line observations, which often only trace the dense gas,
   have mainly revealed cores which are broader than the filament
   \citep{hacar2011, hacar2013, tafalla2015}. Thus, the interpretation
   of core radius is complex and core morphology obviously depends on the tracer
   of observation.\\
   Numerical predictions by \citet{nagasawa1987} showed that
   there are two regimes of the perturbation. One for low line-mass filaments,
   called deformation instability or "sausage" instability, where the
   forming cores bulge out and one for high line-mass, named compressional
   instability, where cores form by compression and thus pinch in. Both
   morphologies exist in simulations throughout the literature
   \citep{gehman1996a, gehman1996b, inutsuka1997, fiege2000}. However, in
   order to determine the morphology of cores it is important to not only
   predict the radius evolution of the core itself, but also the radius
   evolution of the material making up the rest of the filament. For a
   growing perturbation, both evolve simultaneously. We expand on the
   picture by \citet{nagasawa1987} and show an analytical prediction
   for the evolution of the radius ratio.

\section{Basic concepts}
\label{sec:concepts}

   In order to be able to calculate the radial extent of a filament,
   it is necessary to define the underlying density structure. The basic
   hydrostatic, isothermal model predicts a profile which drops off as
   $r^{-4}$ \citep{stodolkiewicz1963, ostriker1964}. Observationally,
   filaments often show a shallower power law exponent of -1.6 to -2.5 at
   large radii \citep{arzoumanian2011, palmeirim2013}. Several
   processes can explain this difference: truncation of the filament radius
   in pressure equilibrium \citep{fischera2012}, magnetic fields
   \citep{fiege2000}, the equation of state \citep{gehman1996a, toci2015}
   or filaments formed by shock interaction \citep{federrath2016}. As the
   physical reason for the observed profile and how it would impact
   the radial stability is still unclear, we use the basic isothermal model.
   In this case the density goes as:
   \begin{equation}
     \rho(r) = \rho_c \left(1+\left(r/H\right)^{2}\right)^{-2}
   \end{equation}
   where $r$ is the cylindrical radius and $\rho_c$ is its central density.
   It has the radial scale height $H$ given by
   \begin{equation}
      H^2 = \frac{2c_s^2}{\pi G \rho_c}
   \end{equation}
   where $c_s$ is the isothermal sound speed and $G$ is the gravitational
   constant. Integrating the density profile to $r \rightarrow \infty$,
   one can calculate the critical line-mass, e.g. the line-mass at which a
   filament is marginally stable, of
   \begin{equation}
      \left(\frac{M}{L}\right)_\text{crit} = \frac{2c_s^2}{G}.
   \end{equation}
   If the line-mass of a filament is above this value, there is no
   hydrostatic solution and the filament will collapse to a spindle.
   If the line-mass is below this value the filament will expand freely
   unless it is bound by an additional outside pressure \citep{nagasawa1987}.
   In this case, the filament follows the hydrostatic equilibrium profile
   until it extends to the radius where the internal pressure matches
   the external pressure. Following \citet{fischera2012}, the integral
   of the density profile then is given by
   \begin{equation}
      \frac{M}{L} = \int_0^R 2 \pi r \rho(r) dr =
      \left(\frac{M}{L}\right)_\text{crit} \left(1+\left(H/R\right)^2\right)^{-1}
      \label{eq:intlm}
   \end{equation}
   and the factor of line-mass to critical line-mass becomes
   \begin{equation}
     f_\text{cyl} = \left.\left(\frac{M}{L}\right)\middle/
     \left(\frac{M}{L}\right)_\text{crit}\right. =
     \left(1+\left(H/R\right)^2\right)^{-1}.
   \end{equation}
   This allows us to derive the filament radius as
   \begin{equation}
      R=H\left(\frac{f_\text{cyl}}{1-f_\text{cyl}}\right)^{1/2}.
      \label{eq:radius}
   \end{equation}
   For a fixed external pressure the scale height is not set by the
   central density but by the ambient pressure via the boundary density
   $\rho_b=p_\text{ext}/c_s^2$. It is related to the central density by
   \begin{equation}
      \rho_b=\rho_c\left(1-f_\text{cyl}\right)^2
      \label{eq:rhoboundary}
   \end{equation}
   and therefore the scale height adjusts as
   \begin{equation}
      H^2 = \frac{2c_s^2}{\pi G \rho_b}\left(1-f_\text{cyl}\right)^2 =
      \frac{2c_s^4}{\pi G p_\text{ext}}\left(1-f_\text{cyl}\right)^2.
   \end{equation}
   Subsequently, the radius has a maximum at $f_\text{cyl}=0.5$ and declines
   to zero as $f_\text{cyl}$ approaches 0 or 1.
   \\
   Linear perturbation analysis introduces a perturbation along the
   filament axis of the form
   \begin{equation}
      \rho(r,z,t)=\rho_0(r)+\rho_1(r,z,t)=
                  \rho_0(r)+\epsilon\rho_0(r)\exp(ikz-i\omega t)
      \label{eq:rhopert}
   \end{equation}
   where $z$ is the filament axis, $\omega=2\pi/\tau$ is the
   perturbation growth rate with $\tau$ being the perturbation
   growth time, $k=2\pi/\lambda$ is the wave vector with $\lambda$
   being the perturbation length scale and $\epsilon$ is the
   perturbation strength. This also leads to a perturbation in velocity,
   pressure and potential of the form:
   \begin{equation}
      q_1(r,z,t) \propto \exp(ikz-i\omega t).
   \end{equation}
   Solving the mass and momentum conservation as well as Laplace's
   equation for the gravitational potential while second order terms
   are ignored, perturbations grow for values of $k$ where the
   solution of the resulting dispersion relation $\omega^2(k)$ is
   smaller than zero. As the perturbation term of \autoref{eq:rhopert}
   does not depend on radius, one can insert it into the definition of
   the line-mass \autoref{eq:intlm} and easily show that the line-mass
   \begin{equation}
      f_\text{cyl}(z,t) = f_0 \left(1+\epsilon\exp\left(ikz-i\omega t\right)\right)
   \end{equation}
   evolves analogous to the density with $f_0$ being the initial line-mass.
   Therefore, the filament radius now depends on the local line-mass at the
   position $z$.

\section{Core morphology}
\label{sec:morphology}

   While \citet{nagasawa1987} already pointed out the two different
   core formation regimes, it is important to look at the dynamical
   evolution of both core and filament radius. As the radius has its maximum
   at half the critical line-mass, it can both grow or shrink
   for an increase or decrease in the local line-mass, depending on the
   initial line-mass. In low line-mass filaments, where the mean line-mass is
   below half the critical value, the growing core will first increase in radius.
   But as soon as its local line-mass exceeds a value of $f_\text{cyl}=0.5$,
   the radius will decrease again. At the same time, in absence of accretion
   the core is fed by filament gas, thus reducing the line-mass and
   the radius of the rest of the filament. Contrarily, in a filament with
   initially high line-mass, where the mean line-mass is above half the
   critical value, the radius of the core decreases as it grows. As mass is
   accreted from the rest of the filament, the overall filament radius will at
   first increase but then also decrease as soon as the local line-mass is
   below a value of $f_\text{cyl}=0.5$.

\begin{figure}
     \centering
     \includegraphics[width=0.9\columnwidth]{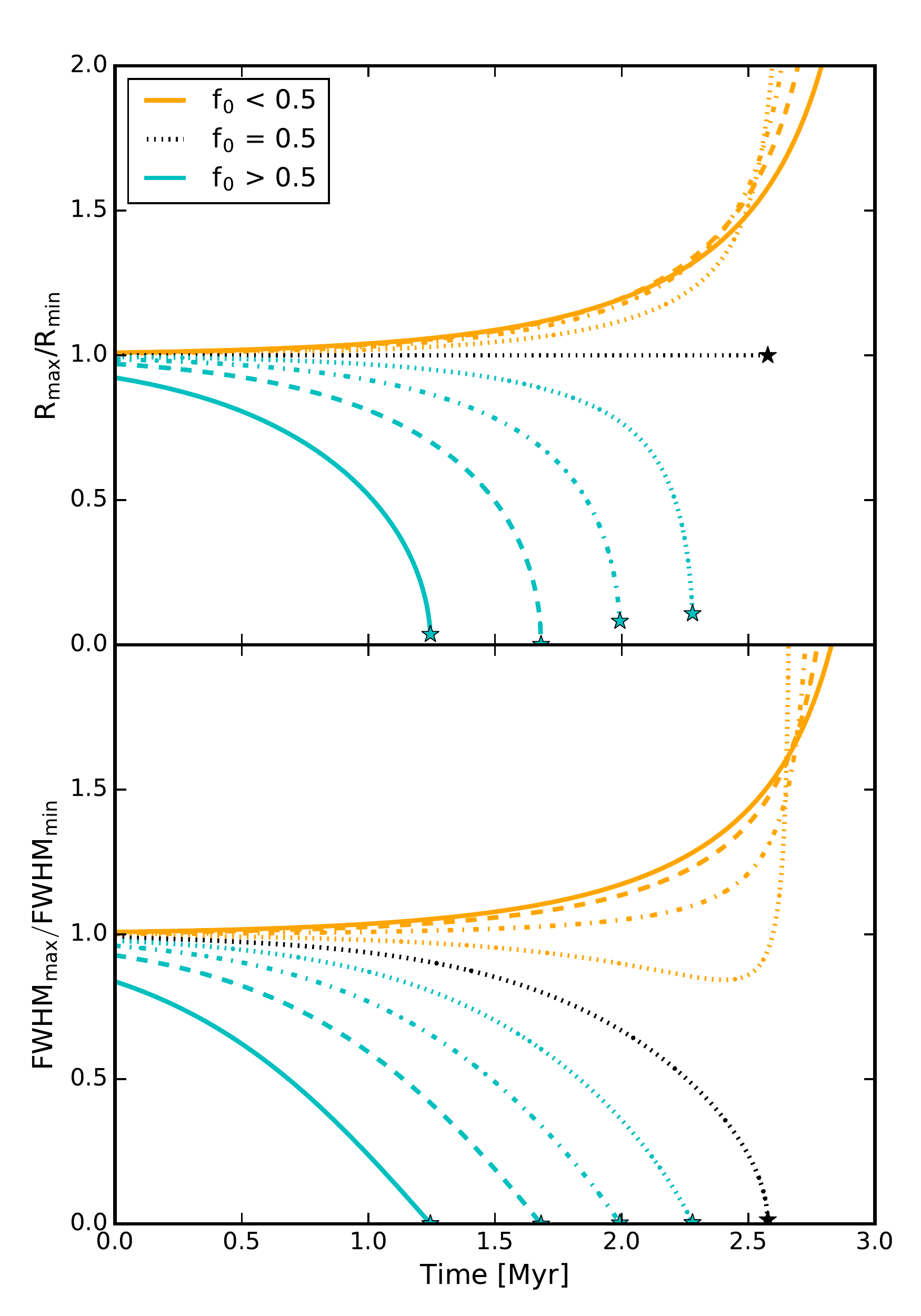}
     \caption{The top figure shows the analytical prediction of the ratio
     between the core radius (at the maximum central density) and filament
     radius (at the minimum central density) for a constant initial central
     density of $\rho_0 = 10^4\ccm$, starting with a perturbation of 1\%.
     The bottom figure shows the same for the FWHM. From top to bottom the
     curves show different values of the line-mass in values of $f_0$,
     starting at 0.1 (solid orange) and incrementally increased by 0.1. The
     stars symbolize the time when the line-mass locally becomes
     supercritical and the core collapses. Note that the radius
     ratio of cores in low line-mass filaments diverges to infinity as all
     the gas of the filament is accreted.}
     \label{fig:rhocons}
   \end{figure}

   In order to determine the core morphology, one has to compare the
   radius $R_\text{max}$ of the slice with the maximum line-mass to the radius
   $R_\text{min}$ of the slice with the minimum line-mass. In a perturbed
   filament both evolve simultaneously and determine how a core appears visually.
   If $R_\text{max} > R_\text{min}$ then the core will bulge out and will be
   broader than the rest of the filament gas. If $R_\text{max} < R_\text{min}$
   the core will be narrower than the filament gas and will pinch in. The radii
   are given by the respective scale height and line-mass as shown in
   \autoref{eq:radius}. The ratio of the two is given by
   \begin{equation}
     \begin{aligned}
       R_\text{max}/R_\text{min} {} & =
       \frac{H_\text{max}\left(f_\text{max}/(1-f_\text{max})\right)^{1/2}}
       {H_\text{min}\left(f_\text{min}/(1-f_\text{min})\right)^{1/2}} = \\
       & = \left(\frac{f_\text{max}(1-f_\text{max})}
       {f_\text{min}(1-f_\text{min})}\right)^{1/2}
      \end{aligned}
   \end{equation}
   with
   \begin{equation}
      f_\text{max} = f_0(1+\epsilon\exp(\omega t)) = f_0 c_+
   \end{equation}
   and
   \begin{equation}
      f_\text{min} = f_0(1-\epsilon\exp(\omega t)) = f_0 c_-.
   \end{equation}
   Note that $c_-= 2 - c_+$. This means that
   \begin{equation}
      R_\text{max}/R_\text{min} =
      \left(\frac{c_+ - f_0 c_+^2}{c_- - f_0 c_-^2}\right)^{1/2}.
   \end{equation}
   Setting this equation equal to 1, one can calculate the line-mass where
   cores stay exactly as broad as the filament to be half the critical
   line-mass. For smaller line-masses the ratio between the radii will at all
   times be larger than one and vice versa. This means that for a filament
   which has a line-mass of less than half the critical value, the decrease
   in radius of the core when it reaches a local line-mass greater than half
   the critical value will always be slower than the overall decrease of
   radius due to the loss of mass in the rest of the filament. Therefore, the
   core will bulge out at all times. The inverse is true for filaments with
   a line-mass above half the critical line-mass, where the core will
   always pinch in. This fact is illustrated in the top panel
   of \autoref{fig:rhocons} where we show the evolution of the ratio between
   core to filament radius over time for a fixed initial central density of
   $10^4 \ccm$ and an initial perturbation strength of 1\%. As the central
   density and the external pressure are not independent of each other, a
   constant central density with a varying initial line-mass means that we
   vary the external pressure from $p_\text{ext}/k_B \sim 10^5\Kccm$ for low
   line-masses to $p_\text{ext}/k_B \sim 10^3\Kccm$ for high line-masses. The
   perturbation growth times are taken from \citet{fischera2012} where we
   assume the perturbation grows on the dominant wavelength. For the same
   initial density the growth time only depends weakly on the line-mass.
   Thus at the same point in time they have evolved by approximately the
   same factor. Note that although the radius in general does depend on the
   initial central density the ratio does not.

   Interestingly, for filaments below half the critical line-mass the radius
   ratio does not depend much on the line-mass itself at a specific point in
   its evolution. As long as the cores have grown by about the same amount we
   do not expect a significant difference in core to filament radius.
   As $f_\text{min}$ goes to zero as soon as the core has accreted nearly all
   material, the radius ratio diverges to infinity. Note that
   \citet{fischera2012} predicted that cores could form an unstable
   Bonnor-Ebert sphere \citep{ebert1955, bonnor1956} depending on
   the perturbation length. We do not include this effect in our
   analytical model but note that it can lead to the collapse of an initially
   broad core and therefore also could lead to a pinched, albeit protostellar
   core.

   \begin{figure}
     \centering
     \includegraphics[width=0.85\columnwidth]{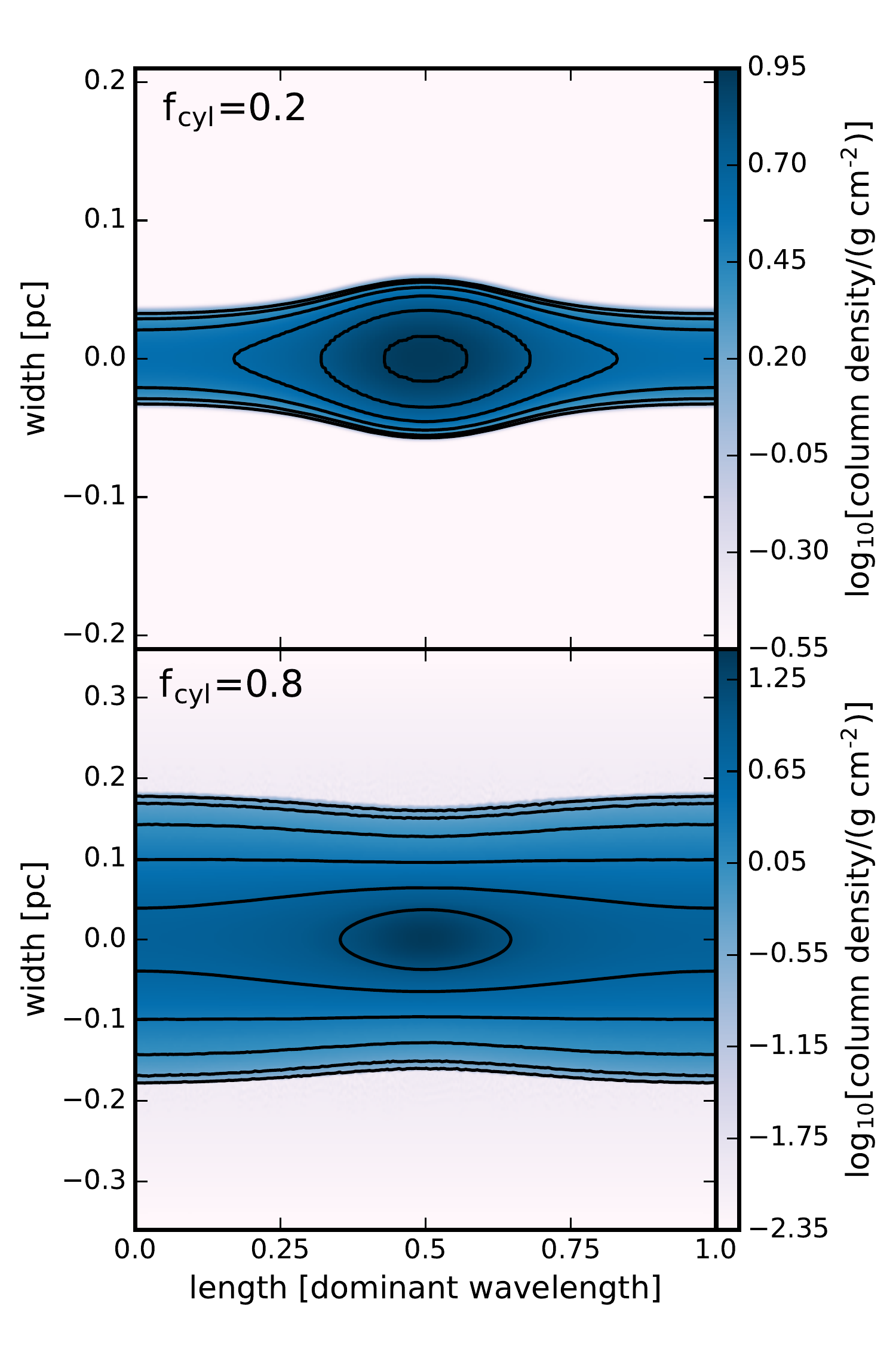}
     \caption{Column density plots of simulated cores forming in filaments
     with different line-masses. A low line-mass filament forms a broad core
     while a core forming in a high line-mass filament causes the radius to
     pinch.}
     \label{fig:pro}
   \end{figure}

   The evolution of cores is significantly different for filaments with a
   line-mass above half the critical value. They tend to evolve much faster
   than their counterparts in low line-mass filaments. As soon as they come
   close to the limit of hydrostatic equilibrium, their radius collapses away
   quite rapidly. This restricts their lifetime and the chance to actually
   observe pinched cores.

   Additionally, we test our predictions by simulating the evolution of cores
   which start with a one per cent perturbation in filaments with a
   central density of $10^4\ccm$ and a line-mass of $f_\text{cyl}=0.2$
   and $0.8$ in order to show the qualitative difference in morphology.
   We use the grid code \textsc{RAMSES} \citep{teyssier2002} to set up boxes
   with the size of the respective dominant perturbation length with periodic
   boundary condition in the filament axis and open boundaries perpendicular
   to the filament. In order to test our prediction independent of
   accretion onto the filaments, they are embedded in a low-density warm
   medium in pressure equilibrium. The simulation
   set-up is similar to that of \citet{heigl2016}. Both results of the
   simulations are shown in projection in \autoref{fig:pro} at the same
   time, shortly before the high line-mass filament collapses. The difference
   in morphology is clearly visible. The core in the low line-mass filament
   is broader then the filament, whereas the core in the high line-mass
   filament causes the radius to decrease. We only see a divergence
   of the Ostriker profile at late times, where both cores profiles become
   softer and closer to the radial dependence of an isothermal sphere.

   \begin{figure}
     \centering
     \includegraphics[width=0.85\columnwidth]{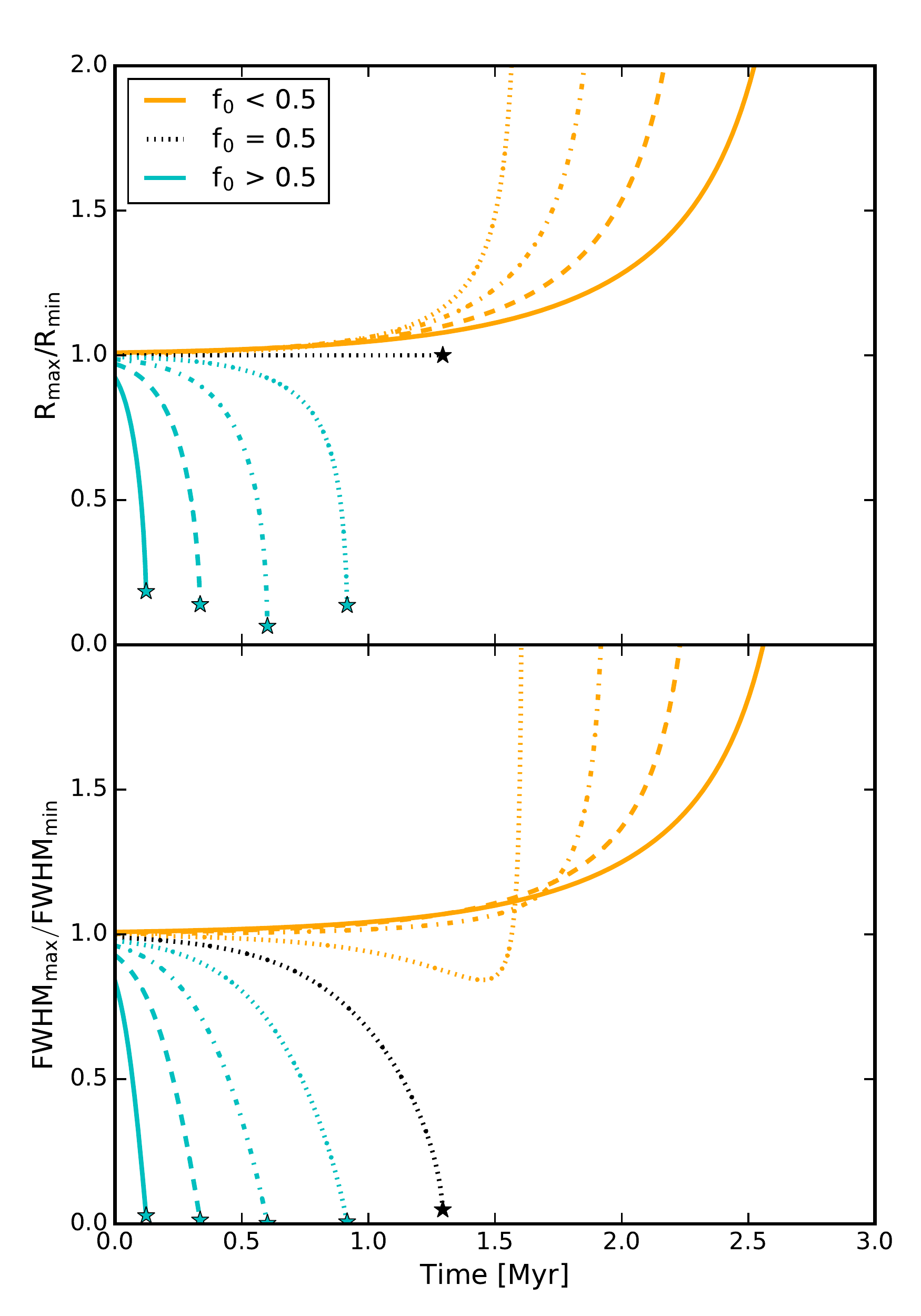}
     \caption{The same as in \autoref{fig:rhocons} but now for a constant
     external pressure of $p_\text{ext}/k_B = 10^5\Kccm$. For varying
     central densities there is a broad spread in perturbation growth time.
     The line properties are the same as in \autoref{fig:rhocons}. Starting
     at $f_0=0.1$ for the curve with the longest perturbation growth time
     (solid orange), $f_0$ increases incrementally by 0.1 going
     to faster perturbation growth times.}
     \label{fig:pcons}
   \end{figure}

   Therefore, our analysis provides observers with a useful tool to determine
   the line-mass a filament, independent of inclination and perturbation
   scale, by identifying the cores. If a core bulges out of the filament,
   the mean line-mass is below 0.5. If a core pinches inwards, the mean
   line-mass is above 0.5.
   A caveat of this method is the way the radius is determined in the
   observations. The filament width is often determined by measuring the
   full-width at half maximum (FWHM) of the radial profile. If the filament
   follows an Ostriker profile, the FWHM is not a perfect tracer of the
   radius and a correction term has to be taken into consideration. This
   correction was derived in \citet{fischera2012} and changes the analytical
   prediction of the radius ratio as shown in the bottom of
   \autoref{fig:rhocons}. In general, using the FWHM will underestimate
   the core radius and therefore introduces a bias to lower radius ratios.
   This effect does not change the predicted curves significantly except for
   cores forming in filaments with $f_\text{cyl}= 0.5$, which would be
   observed as a pinched core.

   We assume that filaments in the same region of a molecular cloud are
   embedded in a constant background pressure. In this case, filaments with
   different line-masses will vary substantially in central density and
   therefore also in the perturbation growth time as it goes as
   $\tau_\text{dom} \sim 1/\sqrt{\rho_c}$. This fact is illustrated in
   \autoref{fig:pcons} where we show the same evolution of the core
   to filament radius ratio as in \autoref{fig:rhocons} under the assumption
   that the cores grow on the dominant timescale now for a constant external
   pressure of $p_\text{ext}/k_B = 10^5\Kccm$. Consequently the central
   density varies from $\rho_c \sim 10^4\ccm$ for low line-masses to
   $\rho_c \sim 10^6\ccm$ for high line-masses. There is an interesting
   dichotomy visible in the state of evolution of the cores. High line-mass
   filaments forming pinched cores evolve much faster than low line-mass
   filaments which form broad cores. This implies that, as long as there is
   not much spread in initial perturbation strength in filaments, a region
   will most likely only contain one population of cores. Either there are
   mainly pinched cores and broad cores will not have had enough time to
   grow or there are mainly broad cores and their high line-mass counterparts
   have already collapsed away and formed stars. Varying the value of the
   external pressure only shifts the evolutionary tracks in
   \autoref{fig:pcons} (higher external pressure imply larger central
   densities and thus faster growth times) and does not change the
   general behavior.

\section{Discussion and conclusions}
\label{sec:discussion}

   The gravitational instability model has several shortcomings.
   The main assumption is that filaments are very idealized cylindrical
   entities where the mean initial line-mass does not vary much along its
   length. Moreover, the filament profile requires a certain timescale to
   adjust to density changes. If the local line-mass varies faster than the
   radius can adjust, a broad core could be embedded in a filament with a
   line-mass larger than half the critical value. There are two processes
   which can lead to a major change in local line-mass on a short timescale.
   On the one hand, mass accretion increases the overall line-mass. Observed
   rates are estimated to be on the order of
   $10-100 \text{ M}_\odot \text{ pc}^{-1} \text{ Myr}^{-1}$
   \citep{palmeirim2013}. On the other hand, a filament will longitudinally
   contract due to self-gravity. In addition, the rapid formation of two
   cores at the ends of the filament seems to be a typical outcome of the
   edge-effect
   \citep{burkert2004}.

   A different equation of state or additional physical contribute to the
   radial stability and can change the morphology of cores. Observed radial
   density profiles are better matched by polytropic indices lower than one
   \citep{toci2015}. As long as there is a maximum radius in dependence of
   the line-mass we still expect a dichotomy in morphology but with the
   division not necessarily at half the critical line-mass.

   Observationally, it is important to not only include the dense gas in order
   to reliably measure both filament and core radius. As the density of the
   outer filament gas is lower than the core gas, the filament radius has to
   be determined with a tracer of low gas density. If only the dense gas is
   observed, e.g. \NH, even cores which are nominally pinched can appear
   broader than the dense gas in the rest of in filament.

   Moreover, projection effects can reduce the length of a filament
   and thus increase the apparent line-mass by a substantial factor. This
   effect is limited by the fact that higher inclined filaments will not
   resemble a filamentary structure.

   Additionally, more cores are observed which are thinner than
   the average widths of star-forming filaments
   \citep{palmeirim2013, marsh2014, roy2014}, indicating that most
   filaments have high line-masses. Nevertheless, higher number statics on
   the local ratio of core-to-filament radius are desirable
   in order to estimate line-masses.

   \noindent All in all, our model allows for the following predictions:
   \begin{itemize}
     \item The morphology of cores embedded in filaments is set by the
           initial line-mass. Filaments with an initial line-mass below
           half the critical value will develop broad cores. Filaments with
           an initial line-mass above half the critical value will develop
           pinched cores.
     \item For filaments which are embedded in the same constant background
           pressure, the perturbation growth times for low and high
           line-masses are drastically different. If all filaments start with
           similar perturbation strengths we expect only one population of
           cores to be present, only pinched cores at early times and broad
           cores at late times.
     \item Using the FWHM to determine the radius underestimates the extent
           of high density regions of the filament and thus underestimates
           the ratio of core to filament radius.
     \item The phase where the radius of pinched cores is significantly
           different from the overall filament radius is very short and
           indicates an imminent collapse due to loss of hydrostatic
           equilibrium.
   \end{itemize}

\section*{Acknowledgements}

   We thank the anonymous referee for the valuable input that greatly
   increased the quality of the paper.
   AB, MG and SH are supported by the priority programme 1573 "Physics of
   the Interstellar Medium" of the German Science Foundation and the Cluster
   of Excellence "Origin and Structure of the Universe". The simulations were
   run using resources of the Leibniz Rechenzentrum (LRZ, Munich; linux
   cluster CoolMUC2).



\bibliographystyle{mnras}
\bibliography{Core.bib}

\begin{thebibliography}{}
\makeatletter
\relax
\def\mn@urlcharsother{\let\do\@makeother \do\$\do\&\do\#\do\^\do\_\do\%\do\~}
\def\mn@doi{\begingroup\mn@urlcharsother \@ifnextchar [ {\mn@doi@}
  {\mn@doi@[]}}
\def\mn@doi@[#1]#2{\def\@tempa{#1}\ifx\@tempa\@empty \href
  {http://dx.doi.org/#2} {doi:#2}\else \href {http://dx.doi.org/#2} {#1}\fi
  \endgroup}
\def\mn@eprint#1#2{\mn@eprint@#1:#2::\@nil}
\def\mn@eprint@arXiv#1{\href {http://arxiv.org/abs/#1} {{\tt arXiv:#1}}}
\def\mn@eprint@dblp#1{\href {http://dblp.uni-trier.de/rec/bibtex/#1.xml}
  {dblp:#1}}
\def\mn@eprint@#1:#2:#3:#4\@nil{\def\@tempa {#1}\def\@tempb {#2}\def\@tempc
  {#3}\ifx \@tempc \@empty \let \@tempc \@tempb \let \@tempb \@tempa \fi \ifx
  \@tempb \@empty \def\@tempb {arXiv}\fi \@ifundefined
  {mn@eprint@\@tempb}{\@tempb:\@tempc}{\expandafter \expandafter \csname
  mn@eprint@\@tempb\endcsname \expandafter{\@tempc}}}

\bibitem[\protect\citeauthoryear{{Andr{\'e}} et~al.,}{{Andr{\'e}}
  et~al.}{2010}]{andre2010}
{Andr{\'e}} P.,  et~al., 2010, \mn@doi [\aap] {10.1051/0004-6361/201014666},
  \href {http://adsabs.harvard.edu/abs/2010A%26A...518L.102A} {518, L102}

\bibitem[\protect\citeauthoryear{{Andr{\'e}}, {Di Francesco}, {Ward-Thompson},
  {Inutsuka}, {Pudritz}  \& {Pineda}}{{Andr{\'e}} et~al.}{2014}]{andre2014}
{Andr{\'e}} P.,  {Di Francesco} J.,  {Ward-Thompson} D.,  {Inutsuka} S.-I.,
  {Pudritz} R.~E.,   {Pineda} J.~E.,  2014, \mn@doi [Protostars and Planets VI]
  {10.2458/azu_uapress_9780816531240-ch002}, \href
  {http://adsabs.harvard.edu/abs/2014prpl.conf...27A} {pp 27--51}

\bibitem[\protect\citeauthoryear{{Arzoumanian} et~al.,}{{Arzoumanian}
  et~al.}{2011}]{arzoumanian2011}
{Arzoumanian} D.,  et~al., 2011, \mn@doi [\aap] {10.1051/0004-6361/201116596},
  \href {http://adsabs.harvard.edu/abs/2011A%26A...529L...6A} {529, L6}

\bibitem[\protect\citeauthoryear{{Arzoumanian}, {Andr{\'e}}, {Peretto}  \&
  {K{\"o}nyves}}{{Arzoumanian} et~al.}{2013}]{arzoumanian2013}
{Arzoumanian} D.,  {Andr{\'e}} P.,  {Peretto} N.,   {K{\"o}nyves} V.,  2013,
  \mn@doi [\aap] {10.1051/0004-6361/201220822}, \href
  {http://adsabs.harvard.edu/abs/2013A%26A...553A.119A} {553, A119}

\bibitem[\protect\citeauthoryear{{Benson} \& {Myers}}{{Benson} \&
  {Myers}}{1989}]{benson1989}
{Benson} P.~J.,  {Myers} P.~C.,  1989, \mn@doi [\apjs] {10.1086/191365}, \href
  {http://adsabs.harvard.edu/abs/1989ApJS...71...89B} {71, 89}

\bibitem[\protect\citeauthoryear{{Bonnor}}{{Bonnor}}{1956}]{bonnor1956}
{Bonnor} W.~B.,  1956, \mn@doi [\mnras] {10.1093/mnras/116.3.351}, \href
  {http://adsabs.harvard.edu/abs/1956MNRAS.116..351B} {116, 351}

\bibitem[\protect\citeauthoryear{{Burkert} \& {Hartmann}}{{Burkert} \&
  {Hartmann}}{2004}]{burkert2004}
{Burkert} A.,  {Hartmann} L.,  2004, \mn@doi [\apj] {10.1086/424895}, \href
  {http://adsabs.harvard.edu/abs/2004ApJ...616..288B} {616, 288}

\bibitem[\protect\citeauthoryear{{Ebert}}{{Ebert}}{1955}]{ebert1955}
{Ebert} R.,  1955, \zap, \href
  {http://adsabs.harvard.edu/abs/1955ZA.....37..217E} {37, 217}

\bibitem[\protect\citeauthoryear{{Federrath}}{{Federrath}}{2016}]{federrath2016}
{Federrath} C.,  2016, \mn@doi [\mnras] {10.1093/mnras/stv2880}, \href
  {http://adsabs.harvard.edu/abs/2016MNRAS.457..375F} {457, 375}

\bibitem[\protect\citeauthoryear{{Fiege} \& {Pudritz}}{{Fiege} \&
  {Pudritz}}{2000}]{fiege2000}
{Fiege} J.~D.,  {Pudritz} R.~E.,  2000, \mn@doi [\mnras]
  {10.1046/j.1365-8711.2000.03067.x}, \href
  {http://adsabs.harvard.edu/abs/2000MNRAS.311..105F} {311, 105}

\bibitem[\protect\citeauthoryear{{Fischera} \& {Martin}}{{Fischera} \&
  {Martin}}{2012}]{fischera2012}
{Fischera} J.,  {Martin} P.~G.,  2012, \mn@doi [\aap]
  {10.1051/0004-6361/201218961}, \href
  {http://adsabs.harvard.edu/abs/2012A%26A...542A..77F} {542, A77}

\bibitem[\protect\citeauthoryear{{Gehman}, {Adams}, {Fatuzzo}  \&
  {Watkins}}{{Gehman} et~al.}{1996a}]{gehman1996a}
{Gehman} C.~S.,  {Adams} F.~C.,  {Fatuzzo} M.,   {Watkins} R.,  1996a, \mn@doi
  [\apj] {10.1086/176766}, \href
  {http://adsabs.harvard.edu/abs/1996ApJ...457..718G} {457, 718}

\bibitem[\protect\citeauthoryear{{Gehman}, {Adams}  \& {Watkins}}{{Gehman}
  et~al.}{1996b}]{gehman1996b}
{Gehman} C.~S.,  {Adams} F.~C.,   {Watkins} R.,  1996b, \mn@doi [\apj]
  {10.1086/178098}, \href {http://adsabs.harvard.edu/abs/1996ApJ...472..673G}
  {472, 673}

\bibitem[\protect\citeauthoryear{{Hacar} \& {Tafalla}}{{Hacar} \&
  {Tafalla}}{2011}]{hacar2011}
{Hacar} A.,  {Tafalla} M.,  2011, \mn@doi [\aap] {10.1051/0004-6361/201117039},
  \href {http://adsabs.harvard.edu/abs/2011A%26A...533A..34H} {533, A34}

\bibitem[\protect\citeauthoryear{{Hacar}, {Tafalla}, {Kauffmann}  \&
  {Kov{\'a}cs}}{{Hacar} et~al.}{2013}]{hacar2013}
{Hacar} A.,  {Tafalla} M.,  {Kauffmann} J.,   {Kov{\'a}cs} A.,  2013, \mn@doi
  [\aap] {10.1051/0004-6361/201220090}, \href
  {http://adsabs.harvard.edu/abs/2013A%26A...554A..55H} {554, A55}

\bibitem[\protect\citeauthoryear{{Heigl}, {Burkert}  \& {Hacar}}{{Heigl}
  et~al.}{2016}]{heigl2016}
{Heigl} S.,  {Burkert} A.,   {Hacar} A.,  2016, \mn@doi [\mnras]
  {10.1093/mnras/stw2271}, \href
  {http://adsabs.harvard.edu/abs/2016MNRAS.463.4301H} {463, 4301}

\bibitem[\protect\citeauthoryear{{Hosseinirad}, {Naficy}, {Abbassi}  \&
  {Roshan}}{{Hosseinirad} et~al.}{2017}]{hosseinirad2017}
{Hosseinirad} M.,  {Naficy} K.,  {Abbassi} S.,   {Roshan} M.,  2017, \mn@doi
  [\mnras] {10.1093/mnras/stw2820}, \href
  {http://adsabs.harvard.edu/abs/2017MNRAS.465.1645H} {465, 1645}

\bibitem[\protect\citeauthoryear{{Inutsuka} \& {Miyama}}{{Inutsuka} \&
  {Miyama}}{1997}]{inutsuka1997}
{Inutsuka} S.-i.,  {Miyama} S.~M.,  1997, \apj, \href
  {http://adsabs.harvard.edu/abs/1997ApJ...480..681I} {480, 681}

\bibitem[\protect\citeauthoryear{{Kirk} et~al.,}{{Kirk}
  et~al.}{2013}]{kirk2013}
{Kirk} J.~M.,  et~al., 2013, \mn@doi [\mnras] {10.1093/mnras/stt561}, \href
  {http://adsabs.harvard.edu/abs/2013MNRAS.432.1424K} {432, 1424}

\bibitem[\protect\citeauthoryear{{Klessen}, {Burkert}  \& {Bate}}{{Klessen}
  et~al.}{1998}]{klessen1998}
{Klessen} R.~S.,  {Burkert} A.,   {Bate} M.~R.,  1998, \mn@doi [\apjl]
  {10.1086/311471}, \href {http://adsabs.harvard.edu/abs/1998ApJ...501L.205K}
  {501, L205}

\bibitem[\protect\citeauthoryear{{Klessen}, {Ballesteros-Paredes},
  {V{\'a}zquez-Semadeni}  \& {Dur{\'a}n-Rojas}}{{Klessen}
  et~al.}{2005}]{klessen2005}
{Klessen} R.~S.,  {Ballesteros-Paredes} J.,  {V{\'a}zquez-Semadeni} E.,
  {Dur{\'a}n-Rojas} C.,  2005, \mn@doi [\apj] {10.1086/427255}, \href
  {http://adsabs.harvard.edu/abs/2005ApJ...620..786K} {620, 786}

\bibitem[\protect\citeauthoryear{{K{\"o}nyves} et~al.,}{{K{\"o}nyves}
  et~al.}{2010}]{koenyves2010}
{K{\"o}nyves} V.,  et~al., 2010, \mn@doi [\aap] {10.1051/0004-6361/201014689},
  \href {http://adsabs.harvard.edu/abs/2010A%26A...518L.106K} {518, L106}

\bibitem[\protect\citeauthoryear{{Larson}}{{Larson}}{1985}]{larson1985}
{Larson} R.~B.,  1985, \mn@doi [\mnras] {10.1093/mnras/214.3.379}, \href
  {http://adsabs.harvard.edu/abs/1985MNRAS.214..379L} {214, 379}

\bibitem[\protect\citeauthoryear{{Marsh} et~al.,}{{Marsh}
  et~al.}{2014}]{marsh2014}
{Marsh} K.~A.,  et~al., 2014, \mn@doi [\mnras] {10.1093/mnras/stu219}, \href
  {http://adsabs.harvard.edu/abs/2014MNRAS.439.3683M} {439, 3683}

\bibitem[\protect\citeauthoryear{{McKee} \& {Ostriker}}{{McKee} \&
  {Ostriker}}{2007}]{mckee2007}
{McKee} C.~F.,  {Ostriker} E.~C.,  2007, \mn@doi [\araa]
  {10.1146/annurev.astro.45.051806.110602}, \href
  {http://adsabs.harvard.edu/abs/2007ARA%26A..45..565M} {45, 565}

\bibitem[\protect\citeauthoryear{{Men'shchikov} et~al.,}{{Men'shchikov}
  et~al.}{2010}]{menshchikov2010}
{Men'shchikov} A.,  et~al., 2010, \mn@doi [\aap] {10.1051/0004-6361/201014668},
  \href {http://adsabs.harvard.edu/abs/2010A%26A...518L.103M} {518, L103}

\bibitem[\protect\citeauthoryear{{Myers}}{{Myers}}{2009}]{myers2009}
{Myers} P.~C.,  2009, \mn@doi [\apj] {10.1088/0004-637X/700/2/1609}, \href
  {http://adsabs.harvard.edu/abs/2009ApJ...700.1609M} {700, 1609}

\bibitem[\protect\citeauthoryear{{Nagasawa}}{{Nagasawa}}{1987}]{nagasawa1987}
{Nagasawa} M.,  1987, \mn@doi [Progress of Theoretical Physics]
  {10.1143/PTP.77.635}, \href
  {http://adsabs.harvard.edu/abs/1987PThPh..77..635N} {77, 635}

\bibitem[\protect\citeauthoryear{{Ostriker}}{{Ostriker}}{1964}]{ostriker1964}
{Ostriker} J.,  1964, \mn@doi [\apj] {10.1086/148005}, \href
  {http://adsabs.harvard.edu/abs/1964ApJ...140.1056O} {140, 1056}

\bibitem[\protect\citeauthoryear{{Padoan}, {Juvela}, {Goodman}  \&
  {Nordlund}}{{Padoan} et~al.}{2001}]{padoan2001}
{Padoan} P.,  {Juvela} M.,  {Goodman} A.~A.,   {Nordlund} {\AA}.,  2001,
  \mn@doi [\apj] {10.1086/320636}, \href
  {http://adsabs.harvard.edu/abs/2001ApJ...553..227P} {553, 227}

\bibitem[\protect\citeauthoryear{{Palmeirim} et~al.,}{{Palmeirim}
  et~al.}{2013}]{palmeirim2013}
{Palmeirim} P.,  et~al., 2013, \mn@doi [\aap] {10.1051/0004-6361/201220500},
  \href {http://adsabs.harvard.edu/abs/2013A%26A...550A..38P} {550, A38}

\bibitem[\protect\citeauthoryear{{Roy} et~al.,}{{Roy} et~al.}{2014}]{roy2014}
{Roy} A.,  et~al., 2014, \mn@doi [\aap] {10.1051/0004-6361/201322236}, \href
  {https://ui.adsabs.harvard.edu/#abs/2014A&A...562A.138R} {562, A138}

\bibitem[\protect\citeauthoryear{{Schneider} \& {Elmegreen}}{{Schneider} \&
  {Elmegreen}}{1979}]{schneider1979}
{Schneider} S.,  {Elmegreen} B.~G.,  1979, \mn@doi [\apjs] {10.1086/190609},
  \href {http://adsabs.harvard.edu/abs/1979ApJS...41...87S} {41, 87}

\bibitem[\protect\citeauthoryear{{Shu}, {Adams}  \& {Lizano}}{{Shu}
  et~al.}{1987}]{shu1987}
{Shu} F.~H.,  {Adams} F.~C.,   {Lizano} S.,  1987, \mn@doi [\araa]
  {10.1146/annurev.aa.25.090187.000323}, \href
  {http://adsabs.harvard.edu/abs/1987ARA%26A..25...23S} {25, 23}

\bibitem[\protect\citeauthoryear{{Stod{\'o}lkiewicz}}{{Stod{\'o}lkiewicz}}{1963}]{stodolkiewicz1963}
{Stod{\'o}lkiewicz} J.~S.,  1963, \actaa, \href
  {http://adsabs.harvard.edu/abs/1963AcA....13...30S} {13, 30}

\bibitem[\protect\citeauthoryear{{Tafalla} \& {Hacar}}{{Tafalla} \&
  {Hacar}}{2015}]{tafalla2015}
{Tafalla} M.,  {Hacar} A.,  2015, \mn@doi [\aap] {10.1051/0004-6361/201424576},
  \href {http://adsabs.harvard.edu/abs/2015A%26A...574A.104T} {574, A104}

\bibitem[\protect\citeauthoryear{{Teyssier}}{{Teyssier}}{2002}]{teyssier2002}
{Teyssier} R.,  2002, \mn@doi [\aap] {10.1051/0004-6361:20011817}, \href
  {http://adsabs.harvard.edu/abs/2002A%26A...385..337T} {385, 337}

\bibitem[\protect\citeauthoryear{{Toci} \& {Galli}}{{Toci} \&
  {Galli}}{2015}]{toci2015}
{Toci} C.,  {Galli} D.,  2015, \mn@doi [\mnras] {10.1093/mnras/stu2168}, \href
  {http://adsabs.harvard.edu/abs/2015MNRAS.446.2110T} {446, 2110}

\bibitem[\protect\citeauthoryear{{Ward-Thompson} et~al.,}{{Ward-Thompson}
  et~al.}{2010}]{ward-thompson2010}
{Ward-Thompson} D.,  et~al., 2010, \mn@doi [\aap]
  {10.1051/0004-6361/201014618}, \href
  {http://adsabs.harvard.edu/abs/2010A%26A...518L..92W} {518, L92}

\makeatother
\end{thebibliography}







\bsp	
\label{lastpage}
\end{document}